\documentclass[final,5p,times,twocolumn]{elsarticle}

\usepackage{amsmath}
\usepackage{amsfonts}
\usepackage{amssymb}
\usepackage{mathrsfs}

\begin{document}

\begin{frontmatter}



\title{Light-matter interaction and Bose-Einstein condensation of light}




\author[label1]{Vivek M. Vyas}
\address[label1]{Raman Research Institute, C. V. Raman Avenue, Sadashivnagar, Bengaluru 560 080, India}
\ead{vivekv@rri.res.in}

\author[label2]{Prasanta K. Panigrahi}
\address[label2]{Indian Institute of Science Education
and Research Kolkata, Mohanpur,
Nadia 741 246, India}

\author[label3,label4]{V. Srinivasan}
\address[label3]{Department of Theoretical Physics, Guindy Campus, University of Madras, Chennai 600 025, India}
\fntext[label4]{VS thanks Prof. Rita John and the Department of Theoretical Physics, University of Madras for the hospitality.}

\begin{abstract}
The atom - electromagnetic field interaction is studied in the Dicke model, wherein a single field mode is interacting with a collection of two level atoms at thermal equilibrium. It is found that in the superradiant phase of the system, wherein the Bose-Einstein condensation of photons takes place, the notion of photon as an elementary electromagnetic excitation ceases to exist. The phase and intensity excitations of the condensate are found to be the true excitations of electromagnetic field. It is found that in this phase, the atom interacts with these excitations in a distinct coherent transition process, apart from the known stimulated emission/absorption and spontaneous emission processes. In the coherent transition it is found that while the atomic state changes in course of the transition process, the state of electromagnetic field remains unaffected. It is found that the transition probability of such coherent transition process is macroscopically large compared to other stimulated emission/absorption and spontaneous emission processes.    
\end{abstract} 


\end{frontmatter}





A careful study of thermodynamics of photons has always lead to wonderful insights into the working of quantum mechanics. Planck discovered the famous radiation formula, which heralded the quantum revolution, while trying to explain the thermal radiation emitted by an ideal black body. Einstein was able to discern about the existence of photons while trying to understand the thermodynamics of atoms interacting with black body radiation \cite{ecg2}. In the process of arriving at a complete quantum derivation of Planck formula, Bose was led to a careful quantum statistical treatment of non-interacting photon gas trapped in a black body cavity, and to the discovery of Bose-Einstein statistics \cite{reichl}. Einstein generalised the treatment of Bose to other kinds of bosons, thereby discovering Bose-Einstein condensation \cite{reichl}.

In this paper the atom-field interaction is studied in a system wherein the (electromagnetic) field of a single cavity mode is interacting with a large collection of two level atoms at thermal equilibrium. Such systems have been extensively studied from different approaches and have attracted many researchers starting from Einstein \cite{reichl}. He showed that a consistent quantum theory compatible with the statistical mechanics of such a system, requires that the elementary excitation of the electromagnetic field be quantised, which later came to be known as photon, and further that the atom-field interaction is summarised by the processes of stimulated absorption/emission and spontaneous emission of photons. In this paper, we show that these atom-field interaction processes are dependent on which thermodynamic phase is realised by the system, and that their properties can get significantly  altered in different phases of the system. The Dicke model, which describes the interaction of single mode with a collection of two level atoms, and which is known to exhibit a second order phase transition, is employed to convey this point. It must be pointed out that such an investigation is topical and of significance, since it is now possible to realise different phases of interacting photon gas  at thermal equilibrium in a laboratory \cite{klaers2010thermalization, klaers2010, vyas}.

It is found that in the normal phase of the Dicke model, all the three processes of radiation emission/absorption are allowed, at par with the usual expectation. In the superradiant phase however, it is discovered that the
notion of photon as an elementary excitation of the electromagnetic field ceases to exist. This occurs because the system realises a broken symmetry ground state in the form of Bose-Einstein condensed of photons. It turns out that the elementary excitations of such a condensate are \emph{not} photons but are phase and intensity excitations of the condensate, known as Goldstone and Higgs mode respectively. The Goldstone mode being zero energy excitations are found to be dynamically uninteresting. Oddly the field-atom interaction is found to comprise of spontaneous Higgs emission and stimulated Higgs emission/absorption processes, and of a distinct \emph{coherent transition process}. It is found that in such a coherent transition process the atom-field interaction takes place in a perfectly coherent manner, so as not to induce
any change in the latter whatsoever. It turns out that transition probability
of such coherent transition is enormously large compared to any kind of Higgs emission/absorption probability and to
the spontaneous photon emission probability of the normal phase. 

In the next section, the phenomenon of Bose-Einstein condensation, notion of elementary excitation and spontaneous symmetry breaking in non-interacting photon gas are studied. The subsequent section is devoted to the study of atom-field interaction, under the purview of Dicke model, in its two different phases. The paper ends with a brief summary of the results and discussion.

\section{Bose-Einstein condensation and spontaneous symmetry breaking in ideal photon gas}\label{bec}

The average number of photons in a black body cavity occupying a mode $\vec{k}$ with energy $\epsilon_{{k}} \quad (\epsilon_{{k}} = |\vec{k}|)$ and temperature $\beta = 1/T$, obeys the Bose-Einstein distribution function \footnote{Throughout the paper we will be working the natural units such that $k_B =1$, $c=1$ and $\hbar=1$.} \cite{reichl}:
\begin{equation}
  n_{k} = \frac{1}{e^{\beta \epsilon_{{k}} } - 1}.  
\end{equation}
Owing to interaction with the `black' walls, the total photon number in the cavity is not fixed. In other words, the photons have a vanishing chemical potential, and at absolute zero the cavity would not have any photons. 

This distribution function was subsequently generalised by Einstein, to take into account other boson systems with non-vanishing chemical potential, which reads \cite{reichl}:
\begin{equation}
  n_{k} = \frac{1}{e^{\beta (\epsilon_{k} - \mu) } - 1}.  
\end{equation}
It was observed that since $1 < e^{\beta \epsilon_{k} }  < \infty$, it implies that the fugacity $z = e^{\beta \mu }$ is bounded $0 < z < 1$. Einstein observed that, below a critical temperature, the zero momentum state can have singular macroscopic occupation as $z \rightarrow 1$:
\begin{equation}
  n_0 = \frac{z}{1 - z} \: {\longrightarrow} \: \infty.
\end{equation}
Occurrence of such a singularity implies a phase transition, leading to a different phase of the boson gas, often called \emph{Bose-Einstein condensation} \cite{reichl}. Experimentally such a condensation phenomenon has been realised in many boson systems, interestingly in a weakly interacting photon gas it was observed only in 2010 \cite{klaers2010}.

Heisenberg quantised electromagnetic field, and showed that a system of non-interacting photons can actually be thought of as an excited state of quantised electromagnetic field. The Hamiltonian for such a system is given by:
\begin{equation} \label{h1}
 H = \int d^{3}x \: \left( \vec{E}^2 + \vec{B}^2 \right) = \sum_{\vec{k}} \: \epsilon_{k} a^{\dagger}_{\vec{k}} a_{\vec{k}}. 
\end{equation}  
The operators $a_{\vec{k}}$ and $a^{\dagger}_{\vec{k}}$ are the photon annihilation and creation operators respectively for mode ${\vec{k}}$ with energy $\epsilon_{k}$, obeying the commutation relations:
\begin{align} \label{cr}
  [a_{\vec{p}}, a^{\dagger}_{\vec{q}} ] = \delta_{\vec{p},\vec{q}}, \: [a_{\vec{p}}, a_{\vec{k}} ] = 0 \: \: \text{and} \: \: [a^{\dagger}_{\vec{p}}, a^{\dagger}_{\vec{q}} ] = 0.
\end{align}
A state of $n$-photons, which is an eigenstate of (photon) number operator $N = \sum_{\vec{k}} a^{\dagger}_{\vec{k}} a_{\vec{k}}$, often called the number state, can be constructed in this framework from the state $| 0 \rangle$, by application of creation operators $a^{\dagger}_{\vec{k}}$ \cite{huang, agarwal}. The $| 0 \rangle$ state is assumed to be the one where there are no photons, often referred to as a no-photon state. This is ensured by demanding that $a_{\vec{k}} | 0 \rangle = 0$, so that $N | 0 \rangle = \sum_{\vec{k}} a^{\dagger}_{\vec{k}} a_{\vec{k}} | 0 \rangle = 0 $.

The treatment of photon gas in the quantised field framework, also often called the second quantised formalism, can be naturally  generalised to study the thermodynamics of such a system.  One can readily use (\ref{h1}) and evaluate the grand partition function:
\begin{equation}
  Z = {\text{Tr} \: e^{- \beta (H - \mu N)}}, 
\end{equation}
from which the Helmholtz free energy $F$ can be found: $F = - \beta \: \text{ln} \: Z$, whose minimisation gives the information about the stable phase of the system \cite{reichl, ume}. The thermal average of any observable $A$ can now be evaluated as:
\begin{equation} \label{tave}
  \langle A \rangle_{\beta} = \frac{1}{Z} \text{Tr} \left( A \: e^{- \beta (H - \mu N)} \right).
\end{equation}
In particular, the distribution function $n_{k}$ can now be understood as \cite{reichl, ume}: 
\begin{equation} \label{n1}
n_{k} = \frac{1}{Z} \text{Tr} \left( a^{\dagger}_{k} a_{k} \: e^{- \beta (H - \mu N)} \right).  
\end{equation}

Consider the system of photon gas at equilibrium inside a black body cavity, where $\mu = 0$. Since $n_{0} \rightarrow 0$ as $\beta \rightarrow \infty$, it turns out that the ground state of the system is no-photon state $| 0 \rangle$. It can be shown that the Hilbert space realised by the system is the one which is spanned by the number states, by evaluating trace in (\ref{n1}) using the number states basis, and arriving at the Bose-Einstein formula \cite{ume2, knight}.

On the otherhand, consider the case when the photon gas system admits a non-zero chemical potential $\mu \neq 0$. This happens for example when the photon gas is at equilibrium in a Fabry-Perot cavity \cite{chiao2000bogoliubov}. In the limit $\beta \rightarrow 0$, the system undergoes Bose-Einstein condensation - macroscopic population of the ground state $n_0 {\rightarrow} \: \infty$, which can not certainly be described by no-photon state, or any other state number state. A careful reflection reveals that the ground state of the system $|vac \rangle$ can not be the one that is annihilated by $a_{\vec{k}}$, instead should be an eigenstate of $a_{\vec{k}}$: $a_{\vec{k}} | vac \rangle = \alpha_{\vec{k}} |vac \rangle$, where $\alpha_{\vec{k}} \propto \delta(\vec{k})$. Thus, in the limit $\beta \rightarrow \infty$, it turns out that $n_0 = \langle vac | a^{\dagger}_0 a_0 | vac \rangle = |\alpha_0|^2 \rightarrow \infty$ \cite{ume}. Such states which are eigenstates of annihilation operator $a_{\vec{k}}$ are known in the literature as \emph{coherent states} \cite{ecg, agarwal}. Thus one sees that the Hilbert space realised by the system is the one where the ground state is the coherent state $| vac \rangle$. 
As in the earlier case one construct a complete orthonormal basis from the coherent ground state by using the shifted creation and annihilation operators $b_{\vec{k}} = a_{\vec{k}} - \alpha_{\vec{k}}$ and $b^{\dagger}_{\vec{k}} = a^{\dagger}_{\vec{k}} - \alpha^{\ast}_{\vec{k}}$, noting that they obey the same commutation relations as (\ref{cr}):  
\begin{align} 
  [b_{\vec{p}}, b^{\dagger}_{\vec{q}} ] = \delta_{\vec{p},\vec{q}}, \: [b_{\vec{p}}, b_{\vec{k}} ] = 0 \: \: \text{and} \: \: [b^{\dagger}_{\vec{p}}, b^{\dagger}_{\vec{q}} ] = 0.
\end{align}
With these relations it is possible to create excited states above the coherent ground state, which is annihilated by $b_{\vec{k}}$: $b_{\vec{k}} | vac \rangle =0$. It can be checked that the excited states thus obtained, for example $| {\vec{k}} \rangle_{\alpha} = b_{\vec{k}}^{\dagger} | vac\rangle$, $| {\vec{p}}, {\vec{q}} \rangle_{\alpha} = b_{\vec{p}}^{\dagger} b_{\vec{q}}^{\dagger}| vac\rangle$ and so on, indeed form a complete orthonormal basis in the Hilbert space realised by the system. It is worth mentioning that, all the vectors that form this Hilbert space are orthogonal to the vectors that can be expressed as linear combination of number states. This can be simply seen by noting that:
\begin{equation} \label{overlap}
\langle 0  | vac \rangle = \langle 0 | \exp \left( {- \sum_{\vec{k}} |\alpha_{\vec{k}}|^2 } \right) \exp \left({- \sum_{\vec{k}} \alpha_{\vec{k}} a^{\dagger}_{\vec{k}} } \right)| 0 \rangle \rightarrow 0,
\end{equation} 
since $\sum_{\vec{k}} |\alpha_{\vec{k}}|^2 \rightarrow \infty$. This discussion shows that, a given physical system with a well defined Hamiltonian, under different thermodynamic conditions, can realise completely different Hilbert spaces, which in general would be orthogonal to one another.

Physically the excited states $| {\vec{k}} \rangle_{\alpha}= b_{\vec{k}}^{\dagger} | vac\rangle$ however \emph{do not} represent a photon (which is represented by $| {\vec{k}} \rangle = a_{\vec{k}}^{\dagger} | 0 \rangle$) but a distinct excitation over the Bose-Einstein condensed ground state. Thus the notion of elementary excitations/particles is not an absolute one and crucially depends on the thermodynamic phase realised by the system. This is analogous to a system of macroscopically large collection of molecules that can exist in gaseous and solid phases. In the gaseous phase, a state of the system is defined by  the average magnitude of energy/momentum of a single molecule and deviations from it. Once the system condenses to form a solid, the notion of motion a single isolated molecule does not make sense owing to the lattice structure. What is sensible however is to think in terms of lattice vibrations - phonons, which are excitations over an ordered ground state of molecular lattice. This shows that the notion of elementary excitation is subjective one, elementary excitation of one phase need not always exist in another phase, for example the motion of an isolated individual molecule in solid phase, and lattice vibration - phonon in the gaseous phase do not make sense.

It was shown by Noether, that in a classical system, existence of continuous symmetry transformations gives rises to conservation laws. This powerful connection between symmetry and conservation law, was subsequently generalised by Dirac, Wigner and Schwinger for quantum systems \cite{finkel}. The result states that, if there exists a unitary transformation $U$ which is being generated by a generator $K$, $U = \exp(- i \theta K)$ ($\theta$ is a real number), is such that the Hamiltonian is invariant under its operation $U^{-1} H U = H$, then the corresponding generator $K$ is conserved\footnote{Throughout the paper we have assumed the Heisenberg picture of time evolution.}: $\frac{dK}{dt} = 0 $. Since under the action of such unitary transformations the averages and amplitudes remain invariant, one says that such transformations are \emph{symmetries} of the system. Interestingly, the Hamiltonian $H = \sum_{\vec{k}} \: \epsilon_{\vec{k}} a^{\dagger}_{\vec{k}} a_{\vec{k}}$ of the photon gas system admits a continuous symmetry generated by number operator $N$. Owing to the commutation relations:
\begin{equation}
  [N, a_{\vec{k}}] = - a_{\vec{k}}, \quad [N, a^{\dagger}_{\vec{k}}] =  a^{\dagger}_{\vec{k}},
\end{equation}
one indeed sees that under action of $U = \exp(-i\theta N)$, the operators $a_{\vec{k}}$ and $a^{\dagger}_{\vec{k}}$ receive a phase: $U^{-1} a_{\vec{k}} U = a_{\vec{k}} e^{-i\theta}$, $U^{-1} a^{\dagger}_{\vec{k}} U = a^{\dagger}_{\vec{k}} e^{i\theta}$. As a result the Hamiltonian is invariant under the action of unitary transformation, implying that $U = \exp(- i \theta N)$ is indeed the symmetry of the system. The action of $U$ on a given (normalised) state $|a \rangle$ transforms it to $|a'\rangle$ such that: $|\langle a| a' \rangle|^2 \leq 1 $ while preserving it's norm: $\langle a | a \rangle = \langle a' | a' \rangle = 1$. When the system realises $| 0 \rangle$ as the ground state, its overlap with the transformed ground state $U | 0 \rangle$ is:
\begin{equation}
  \langle 0 | U | 0 \rangle = \langle 0 | e^{- i \theta N} | 0 \rangle = 1,
\end{equation} 
implying that the transformed ground state is same as $|0\rangle$. Thus one says that the ground state $| 0\rangle$ is invariant under action of symmetry transformation $U$, akin to the Hamiltonian \cite{ghk}. This is in sharp contrast, in the case  when the system realises the coherent state $| vac \rangle$ as the ground state. In such a case, the overlap between $| vac \rangle$ and $U | vac \rangle$ is actually divergent:  
\begin{equation}
  \langle vac | U | vac \rangle = \langle vac | \sum_{m=0}^{\infty} \frac{1}{m!}(- i \theta N)^{m} | vac \rangle \rightarrow \infty,
\end{equation}
since $\langle vac | N | vac \rangle$ diverges\footnote{It can be rigorously shown that $N$ is not a well defined operator in this Hilbert space \cite{ghk}.}. This is a scenario where one sees that the transformation $U$ is a symmetry of the system, in the sense that Hamiltonian is invariant, however its action is not well defined on the ground state. In general, whenever a continuous symmetry of the Hamiltonian is not respected by the ground state, it is said that the symmetry is \emph{spontaneously broken} \cite{ghk}. It is evident that the formation of Bose-Einstein condensate has lead to spontaneous breaking of phase symmetry generated by $e^{- i \theta N}$ \cite{ume, ghk}.

From the above discussion it becomes clear that, in the system of non-interacting photon gas, the physics of Bose-Einstein condensation, the notion of photons are elementary excitation, and that of spontaneous symmetry breaking is intimately related to realisation of a particular ground state. In the next section  we shall show that this also holds when the photons are interacting with a large collection of atoms.

\section{Bose-Einstein condensation and spontaneous symmetry breaking in photon gas system interacting with atoms}

The study of electromagnetic radiation interacting with a collection of atoms at thermal equilibrium has a long history. Using the ideas of the old quantum theory, Einstein in 1916, studied the statistical mechanics of such a system, and treated the electromagnetic field as a gas of photons interacting with two fixed levels of atoms at thermal equilibrium \cite{reichl, ecg2}. It was proposed that the atom-field interaction can be summarised by three possible processes: a) absorption of photon by an atom, b) spontaneous emission of photon by an atom, c) stimulated emission of a photon from an atom. Using the framework of quantised  electromagnetic field, Dirac gave a modern derivation of this result of Einstein \cite{dirac}. In what follows, we shall consider a simple exactly soluble model called the Dicke model for understanding the physics of atom-field interaction \cite{dicke}. The model essentially consists of a single cavity field mode interacting with a macroscopically large collection of atoms. The field is assumed to couple to only two fixed energy levels of the atoms, and all the other atomic degrees of freedom are assumed to be irrelevant. This model has been extensively studied from various point of views \cite{hepp, wang, gross, and, popov}.

The Hamiltonian for the Dicke model reads:
\begin{equation} \label{dicke}
\mathrm{H} = a^{\dagger}a + \sum_{i=1}^{N} \epsilon S^{i}_z + \frac{\lambda}{\sqrt{N}} \sum_{i=1}^{N} \left( a S^{i}_{+}  + a^{\dagger} S^{i}_{-} \right).
\end{equation}
Here, $a$ and $a^{\dagger}$ stand respectively for annihilation and creation operator for the single photon mode, which is interacting with the two level atom. Since the Hilbert space of a two level atom is same as that of spin space of a half spin particle, it is often convenient to express the atom dynamics in the language of the latter. The ground state of $i^{th}$ atom is thus identified with the down spin state $|m=-\frac{1}{2}\rangle_i$, whereas the excited state is identified with the up spin state $|m=\frac{1}{2} \rangle_i$. The Hamiltonian of the atom, with $\epsilon$ energy spacing between the two levels, is simply $\epsilon S^{i}_z$. The creation and annihilation operators for the atom, analogous to the photon mode, are given respectively by $S_{+}^{i}$ and $S_{-}^{i}$, so that $|\frac{1}{2}\rangle_i = S_{+}^{i} |-\frac{1}{2}\rangle_i$ and $|-\frac{1}{2}\rangle_i = S_{-}^{i} |\frac{1}{2}\rangle_i$. The spin operators by construction obey the angular momentum algebra:
\begin{equation}
[S_{+}^{i}, S_{-}^{i}] = 2 S_{z}^{i}, \quad [S_{z}^{i}, S_{\pm}^{i}] = \pm S_{\pm}^{i}.
\end{equation}

Interestingly the Dicke model defined by the Hamiltonian (\ref{dicke}) is invariant under a continuous symmetry transformation $U = e^{i \theta K}$, where $K = a^{\dagger}a + \sum_{i=1}^{N} S^{i}_z$. This can be readily seen since:
\begin{align}
&U^{\dagger} a U = e^{i \theta} a, \quad U^{\dagger} a^\dagger U = e^{-i \theta} a^\dagger,\\
&U^{\dagger} \left( \sum_{i=1}^{N} S_{\pm}^{i} \right) U = e^{\mp i \theta} \sum_{i=1}^{N} S_{\pm}^{i}. 
\end{align} 
This invariance of $H$: $U^{\dagger} H U = H$ implies that $K$, which is the sum of number of photons and number of excited atoms, is a conserved quantity under time evolution $[H, K] = 0$. Thus all the energy eigenstates - stationary states are also eigenstates of $K$, and hence can be labelled by two quantum numbers $E$ and $k$, corresponding to $H$ and $K$ respectively. This fact has an important consequence when one considers the possibility of transition from an initial stationary state $|i\rangle$ to a final stationary state $|f\rangle$. The initial and final states need to have same energy, and also same 
$k$ quantum number, essentially giving rise to a selection rule \cite{dirac, ecg2}:
\begin{equation} \label{selec}
\Delta k = \Delta m + \Delta n = 0.
\end{equation}

Inorder to study radiation emission phenomenon in this model, one is required to calculate the transition probability (also called radiation emission rate)  $T_{i \rightarrow f}$:
\begin{equation} \label{tp}
T_{i \rightarrow f} = 2 \pi |M_{fi}|^2 \times \delta(E_{f} - E_{i}),
\end{equation} 
from an initial state $| i \rangle$ with energy $E_i$ to final state $| f \rangle$ with energy $E_f$, with the transition matrix element $M_{fi} = \langle f | H_{int} | i \rangle$ \footnote{The delta function $\delta(E_{f} - E_{i})$ in expression (\ref{tp}) appears inorder to ensure energy conservation. However in most experimental situations the initial and final states themselves are found to have a finite width, owing to which the delta function needs to be replaced by say a Lorentzian with a finite width \cite{agarwal}.}. Since the total spin operators $S_+ = \sum_{i=1}^{N} S^{i}_{+}$ and $S_{-} = \sum_{i=1}^{N} S^{i}_{-}$ appear in the interaction Hamiltonian, one may work in a basis which is spanned by eigenstates of $S_z$, total $S^2$ and photon number $a^{\dagger}a$. The basis consists of states denoted by quantum numbers $j$, $m$ and $n$, such that:
\begin{align}
&S^2 |j,m;n \rangle = j(j+1) |j,m;n \rangle,\\ 
&S_z |j,m;n \rangle = m |j,m;n \rangle, \\
&a^{\dagger}a |j,m;n \rangle = n |j,m;n \rangle. 
\end{align}
Assuming that $N$ is odd, implies that $j$ can take half integer values from $1/2$ till $N/2$, $m$ can take half integer values from -$j$ to $j$, and $n$ can be zero or a positive integer. The transition matrix element for emission of radiation from an initial state $|j,m+1\rangle |n\rangle$ to a final state   
$|j,m\rangle |n+1\rangle$, which respects the selection rule (\ref{selec}) reads:
\begin{align} \label{amp}
M_{fi}&= \frac{\lambda}{\sqrt{N}} \langle n+1 | \langle j,m| a^\dagger S_{-} |j,m+1\rangle |n\rangle\\ \label{ampn}
&= \frac{\lambda}{\sqrt{N}} \times \sqrt{n+1} \times \sqrt{(j-m)(j+m+1)}.
\end{align}
This expression gives the amplitude for both: i) \emph{stimulated emission/absorption process} when $n \neq 0$, and ii) \emph{spontaneous emission process} when $n=0$, from the decay of initial atomic state $|j,m+1\rangle$ to $|j,m\rangle$. The derivation of this formula closely follows that of Dirac, and captures the essence of Einstein's treatment of photon-atom interaction \cite{dirac}.

In the case, when $j=1/2$, $m+1=1/2$ and $n=0$, one obtains the {spontaneous emission amplitude} due a single excited atom as: $M^0_{fi} = \frac{\lambda}{\sqrt{N}}$, for which the transition probability reads:
\begin{equation}
T_{0} = 2 \pi \times \frac{\lambda^2}{N} \times \delta(E_{f} - E_{i}).
\end{equation} 
The spontaneous emission amplitude when all the atoms are excited, so that $j=N/2$ and $m+1=N/2$, turns out to be $\lambda$. The transition probability $T$ is hence enhanced by a factor of $N$:
\begin{equation} \label{tpe}
T = 2 \pi \times \lambda^2 \times \delta(E_{f} - E_{i}) = N T_0. 
\end{equation}  

Dicke showed that a further enhancement by factor of $N$ occurs when the initial atomic state is such that $j=N/2$ and $m \sim 0$ \cite{dicke}. The spontaneous emission amplitude for such a decay is $\frac{\lambda}{\sqrt{N}} \sqrt{\frac{N}{2} (\frac{N}{2} + 1)}$, so that the transition probability reads:
\begin{equation} \label{tpsr}
T_{super} = \frac{N}{2} (\frac{N}{2} + 1) T_0 \approx N^2 T_0. \quad (\text{when $N$ is large}) 
\end{equation}
This phenomenon of enhancement of spontaneous emission rate of a collection of $N$ two level atoms is called \emph{Dicke superradiance}, and the atomic states with $j=N/2, m \sim 0$ are often called the Dicke (superradiant) states \cite{dicke, gross}. This phenomenon has been experimentally observed in many different systems \cite{skri, mly, gross, dasp} and lately has been a subject of interest \cite{bhatti, oppel, agarwal, yi}.

Starting from Hepp and Lieb, this model has been studied under thermal equilibrium condition using several approaches \cite{hepp, wang, popov}. It is found that the model exhibits a second order phase transition, when $\lambda^2 > \epsilon$, with the critical temperature $\beta_c$ given by:
\begin{equation} 
\frac{\epsilon}{\lambda^2} = \tanh (\frac{\epsilon \beta_c}{2}). 
\end{equation}
In the normal phase, above the critical temperature $\beta < \beta_c$, it is found that the photon number thermal average vanishes \cite{wang}:
\begin{equation}
\langle a^{\dagger}a \rangle_\beta = 0. 
\end{equation}
Whereas in the so called \emph{superradiant phase} with $\beta > \beta_c$, it is seen that the photon number thermal average diverges \cite{wang}:
\begin{equation} \label{pave}
\langle a^{\dagger}a \rangle_\beta = (\lambda^2 \sigma^2 - \frac{\epsilon^2}{4 \lambda^2}) \times N,  
\end{equation}
where $\sigma$ solves $2 \sigma = \tanh(\beta \lambda^2 \sigma)$. 

From the discussion of Section (\ref{bec}), one immediately infers that the Dicke model in the normal phase realises no-photon state as the ground state: $a | 0 \rangle = 0$. Consequently the phase symmetry generated by $K$ operator is also respected by the ground state $K | 0 \rangle = 0$. This implies that $K$ is a well defined operator on the whole of Hilbert space, which in turn means that the selection rule (\ref{selec}) holds, as also the calculations pertaining to transition probabilities, from (\ref{amp}) to (\ref{tpsr}). In this phase since the photon number operator $a^\dagger a$ is well defined, its eigenstates $|n\rangle$ which represent definite number of photons, are well defined entities and can be used as a basis set.  

On the other hand, in the superradiant phase of the model, one finds that photon number (thermal) average is divergent, implying that the ground state realised by the system is no longer an exact photon number state, rather is a \emph{Bose-Einstein condensate of photons} as also a coherent state: $a | \alpha \rangle = \alpha | \alpha \rangle$, such that $\alpha^{\ast} \alpha =  \frac{N}{4} \left( \lambda^2  - \frac{\epsilon^2}{\lambda^2}\right)$. This immediately follows from (\ref{pave}) by noting that $\langle a^{\dagger}a \rangle_\beta = \mathrm{Tr} \left( e^{-\beta H} a^{\dagger}a \right)$ tends to ground state average $\langle  a^{\dagger}a \rangle_{ground}$ as ${\beta \rightarrow \infty}$. This clearly means that $\langle 0 | \alpha \rangle \rightarrow \infty$ as $N \rightarrow \infty$, showing that the ground state of the system in this phase is orthogonal to the ground state - no-photon state - of the normal phase. Interestingly one sees that the phase symmetry generated by $K$ operator is not respected by the ground state, since $\langle \alpha | K | \alpha \rangle \propto N $ is ill-defined in the thermodynamic limit $N \rightarrow \infty$. Thus in the superradiant phase of Dicke model the phase symmetry is spontaneously broken. The fact that $K$ is not a well defined operator on the Hilbert space, which is realised by the model in this phase, implies that it's eigenvalues can no longer be used to distinguish two degenerate eigenstates of $H$. It also means that the selection rule (\ref{selec}) is no longer meaningful, as also the calculations of transition probabilities following (\ref{amp}). More importantly since the photon number operator $a^{\dagger}a$ itself becomes ill-defined, it is not meaningful to talk about its eigenstates which represent a definite number of photons. Thus the physical state of the electromagnetic field in the cavity in this phase, can not be understood using the notion of photon(s). Such a notion only makes sense for a given state so long as photon number average $\langle a^{\dagger}a \rangle$, photon number variance $\langle (a^\dagger a)^2 \rangle - \langle a^{\dagger}a \rangle^2$, and other higher order moments are well defined.

Thus inorder to understand atom-field dynamics in this phase, it is essential to find the nature of the excited states of the electromagnetic field above the broken symmetry ground state. It is well known that whenever continuous symmetries are spontaneously broken, Goldstone modes naturally appear \cite{ghk}. Magnons in ferromagnets and Bogoliubov sound modes in superfluids are well known examples of Goldstone modes \cite{huang, ume, ume2}. In case of Dicke model it has been elegantly shown in Ref. \cite{yi,pimentel} that there are two kinds of electromagnetic excitations in this phase: a) Goldstone modes, and b) Higgs modes. The Goldstone modes owing to their very nature are zero energy excitations, whereas the Higgs modes have an energy $E_{H} = \sqrt{1 + 2 \epsilon + \lambda^4}$. The operators $\chi$ and $b$ corresponding respectively to Goldstone and Higgs modes are, to the leading order, related to operator $a$ as:
\begin{align}
a = e^{i \chi} \left(\alpha + b\right).
\end{align}
Being zero energy mode, Goldstone modes are dynamically unimportant, since their presence or absence in a transition process goes unnoticed. Hence in what follows we shall set $e^{i \chi}$ to be unity. 

From the discussion in Section (\ref{bec}) it follows that the excited states containing Higgs modes, which form a complete orthonormal basis set, can be constructed from the application of shifted creation/annihilation operators $b = a - \alpha, \; b^\dagger = a^\dagger - \alpha^\ast$, starting from the ground state $|\alpha \rangle$. By noting that: $b | \alpha \rangle = 0$, the first excited state containing one Higgs particle can be constructed simply using $b^\dagger$: $|1\rangle_\alpha = b^\dagger |\alpha \rangle$. Infact one can obtain $(m+1)^{th}$ excited state from $m^{th}$ by application of $b^\dagger$:
\begin{equation}
b^\dagger |m \rangle_\alpha = \sqrt{m+1} |m+1 \rangle_\alpha
\end{equation}    
while respecting the orthonormality condition: $_{\alpha}\langle m | n\rangle_\alpha = \delta_{m,n}$. In the absence of any Higgs excitation, the second order coherence function \cite{ecg, agarwal}: 
\begin{equation}
g^2(0) = \frac{\langle (a^\dagger a)^2 \rangle - \langle a^\dagger a \rangle}{\langle a^\dagger a \rangle^2},
\end{equation}
equals to unity. Whereas in the presence of a few Higgs particles the system behaves almost like a Bose-Einstein condensed state albeit with thermal noise. This is seen by noting that the coherence function for $m^{th}$ excited state is:
\begin{equation}
g^2(0) = 1 + \frac{2m }{m + \alpha^\ast \alpha} - \frac{(2m+1) m }{(m + \alpha^\ast \alpha)^2}, 
\end{equation}   
which is close to unity when $m \ll N$. However for higher excited states with such that $m \sim N$, the system behaves like a thermal state \cite{agarwal}. Thus one sees that in the broken symmetry phase, the true excitations of the electromagnetic field are not photons created by $a^\dagger$, but are Higgs modes created by $b^\dagger$.
These Higgs modes can be thought of as intensity fluctuations over the Bose-Einstein condensed ground state of photons, whereas the Goldstone modes as phase fluctuations. 

Now that one has a clear idea about the true electromagnetic excitations of the system in this phase, one would like to examine the interaction of these excitations with the atoms. Considering that the system was initially in a state $|j,m+1 \rangle |n \rangle_\alpha$, and decays to $|j,m \rangle |n' \rangle_\alpha$, the transition matrix element for the process can be written as:
\begin{align}
M_{fi} &= {\frac{\lambda}{\sqrt{N}}}    {_\alpha}\langle n' | \langle j,m| S_{-} (b^\dagger + \alpha^\ast) |j,m+1\rangle |n\rangle_\alpha, \\ \label{final}
&= \frac{\lambda}{\sqrt{N}} \times \sqrt{(j-m)(j+m+1)} \times \left( \delta_{n+1,n'} \sqrt{n+1} + \alpha^\ast \delta_{n,n'} \right).
\end{align}
This expression is almost similar to that obtained for the normal phase (\ref{ampn}), except the last term that contains $\alpha^\ast$. The term containing $\delta_{n+1,n'}$ to captures: a) spontaneous emission of Higgs particle by setting $n=0$, and b) stimulated emission of Higgs particle by setting $n \neq 0$.

The last term indicates a process with very large amplitude (since $|\alpha| \propto \sqrt{N}$) wherein an atomic transition takes place without any Higgs emission/absorption since $n=n'$. Intriguingly, this term depicts a transition wherein the initial and final states of the electromagnetic field are \emph{exactly} the same, the transition process does not seem to alter them at all. This transition may be referred to as a \emph{coherent transition}, as opposed to spontaneous and stimulated transition which alter the state of the electromagnetic field in the cavity.  

At first sight this may seem absurd, however a careful reflection reveals that in absence of a selection rule of the type (\ref{selec}) such a transition is not forbidden. Owing to the $\sqrt{N}$ factor, this term contributes predominantly to the matrix element, so that the other contributions can be ignored in large $N$ limit. Evidently this transition owes its origin to Bose-Einstein condensation of photons, and is absent in the normal phase where $\alpha = 0$. Interestingly the transition matrix element is independent of the choice of initial/final state $|n\rangle_\alpha$. By setting $j=1/2$, $m+1=1/2$ in (\ref{final}), one gets the transition matrix element $M^\alpha_{fi} = \frac{\lambda}{\sqrt{N}} \times \alpha^\ast$, depicting interaction of a single atom with the field in state $|n\rangle_\alpha$, for any $n$. The transition probability correspondingly reads:
\begin{align}
T^\alpha_{i \rightarrow f} &= 2 \pi \times \frac{\lambda^2}{N} \times (\alpha^\ast \alpha) \times \delta(E_{f} - E_{i})\\
&= (\alpha^\ast \alpha) \times T_0\\
&= \frac{1}{4}(\lambda^2 - \frac{\epsilon^2}{\lambda^2}) \times N \times T_0.
\end{align}
This result shows that in the broken symmetry phase, the probability of a \emph{coherent transition} for a single atom is enormously large, as compared to the spontaneous transition probability of an atom (in the normal phase). Such a coherent transition as is evident only in the broken symmetry phase of the system, and is completely different from the Dicke superradiance enhancement depicted in (\ref{tpsr}). Furthermore unlike Dicke superradiance enhancement, which is dependent on the choice of the initial atomic state, the enhancement described here holds identically for all the Higgs states in the Hilbert space.

\section{Conclusion}

By carefully studying the Dicke model, an exactly soluble model depicting interaction of single radiation mode with a large collection of two level atoms, it is found that the light-atom interaction depends on the thermodynamics of the system, and can be significantly different in various phases realised by the system. It is shown that in the superradiant phase of the Dicke model, owing to spontaneous breaking of phase symmetry, the notion of photons as well defined excitations of the electromagnetic field ceases to exist. The fact that the ground state of the system in such a case itself is a Bose-Einstein condensate of photons, implies that the true excitations are phase and intensity fluctuations of the condensate. It is found that the phase fluctuations correspond to the zero energy Goldstone mode excitation which naturally occur in a broken symmetry phase, being zero energy excitations they are non-dynamical and hence do not contribute to field-atom interaction. The intensity fluctuations on the other hand are finite energy ones and are referred to as Higgs modes. Interestingly it is found that the field-atom interaction consists not only of spontaneous Higgs emission and stimulated Higgs emission/absorption processes, but also of a coherent transition process, in which case the atom-field interaction takes place in a perfectly coherent manner, so as not to induce any change in the latter whatsoever. It turns out that transition probability of such coherent transition is macroscopically large compared to spontaneous Higgs emission and stimulated Higgs emission/absorption processes and to the spontaneous photon emission process of the normal phase.  
   
A discerning reader may wonder whether this result holds for other atom-photon interacting systems, or is just a singular feature of Dicke model. Inorder address this question, two generalisations of Dicke model
wherein the atoms are three and four level (equispaced ladder type) system coupled with a single photon mode were studied. It was found that both these models also admit a phase transition akin to the Dicke model, with the broken symmetry phase realising the coherent ground state. A careful study showed that in these two cases also the essential results obtained for the Dicke model hold in toto. In fact a careful reflection reveals that these results would also naturally hold for any atom-field interacting system, involving a minimal coupling of field to the atomic electrons, which admits a phase transition, like the Dicke model, leading to a coherent ground state. This can be seen from the fact that the dominant interaction between atoms and the field is governed by the term \cite{sak}: $H_{int} \propto \int d^{3}x \; \vec{p} \cdot \vec{A}(\vec{r},t)$, which gives rise to the transition matrix element: $\langle f | \vec{p} \cdot \vec{A}(\vec{r},t) | i \rangle$. If the system realises a thermodynamic phase such that the ground state is a coherent state $|vac\rangle$ such that: $a_{\vec{k}} | vac \rangle = \alpha_{\vec{k}} | vac \rangle$ ($\alpha$ is a complex function of $\vec{k}$), then using the mode expansion of vector field $\vec{A}(\vec{r},t)$ \cite{agarwal,sak} one immediately sees that the aforementioned conclusions, obtained for the Dicke model regarding radiation emission, will also hold in such a system as well. Apart from the atom-field interacting systems, these results will have important implications in various condensed matter systems, for example the ones involving quantum dots, wherein Dicke superradiance phenomenon has been well studied \cite{r1,r2,r3}.

This work summarily generalises the existing view of atom-field interaction, which was originally put forth by Einstein \cite{dirac,sak,ein}, wherein a possibility of a thermodynamic phase transition was not taken into consideration. 
Since the Bose-Einstein condensation of photons has been realised in experiments, it is hoped that the coherent transition process and the Higgs absorption/emission, as depicted in this work, can be experimentally confirmed.



%

\end{document}